\documentclass[preprint,aps,nofootinbib,showpacs,12pt]{revtex4}

\usepackage{graphicx}
\usepackage{graphics}
\usepackage{epsfig}
\usepackage{color}

\begin{document}\title{{\bf\Large Effects of the Regularization on the 
Restoration of Chiral and Axial Symmetries}}
\author{P. Costa}
\email{pcosta@teor.fis.uc.pt}
\affiliation{Departamento de
F\'{\i}sica, Universidade de Coimbra, P-3004-516 Coimbra, Portugal}
\author{M. C. Ruivo}
\email{maria@teor.fis.uc.pt} \affiliation{Departamento de
F\'{\i}sica, Universidade de Coimbra, P-3004-516 Coimbra, Portugal}
\author{C. A. de Sousa}
\email{celia@teor.fis.uc.pt} \affiliation{Departamento de
F\'{\i}sica, Universidade de Coimbra, P-3004-516 Coimbra, Portugal}

\date{\today}


\begin{abstract}
The effects  of a  type of regularization for  finite temperatures on the restoration 
of chiral and axial symmetries are investigated within the SU(3) Nambu-Jona-Lasinio model.
The regularization consists in using an infinite cutoff in the integrals that are
convergent at finite temperature, a procedure that allows one to take into account the
effects of high momentum quarks  at high temperatures. It is found that the critical
temperature for the phase transition is closer to lattice results than the one obtained
with the conventional regularization, and the   restoration of chiral and axial
symmetries, signaled by the behavior of several observables, occurs simultaneously and at
a higher temperature.  The restoration of the axial symmetry appears as a natural
consequence of the full recovering of the   chiral symmetry  that was dynamically broken. 
By using an additional ansatz that simulates instanton suppression effects, by means of  
a convenient temperature dependence of the anomaly coefficient, we found that the 
restoration of U(2) symmetry is shifted to lower values, but  the dominant effect at 
high temperatures comes from the new regularization that enhances  the decrease of 
quark condensates, especially in the strange sector.
\end{abstract}
\pacs{11.10.Wx, 11.30.Rd, 14.40.Aq}
\maketitle


\section{Introduction}

Although the studies on QCD thermodynamics  have  contributed to the improvement of our
understanding of the QCD phase diagram, many challenging questions remain open. In this
concern, microscopic and phenomenological models  have played a meaningful role and are
expected  to clarify various problems in the future. Phase transitions associated to
deconfinement and restoration of chiral and axial U$_A$(1) symmetries are expected to occur
at high density and/or temperature.  A question that has attracted a lot of attention is
whether  these phase transitions take place simultaneously or not and which observables
could signal their occurrence.

As it is well known,  the Nambu-Jona-Lasinio (NJL) model  has the drawback of being  non
renormalizable, its action containing  ultraviolet divergences that  should be
regularized.  Different types of regularizations may be found in the literature
\cite{Ripka1,Ripka2}, and the sensitivity of different observables to the type of
regularization or value of the cutoff has been discussed \cite{Jaminon, Greiner2}. In the
NJL model, the cutoff used to regularize the quark loop term  is, in general, lower than
1GeV, which limits the domain of applicability  of the model. Several unpleasant features
of the model are due to the fact that the number of levels of the Fermi sea occupied is
restricted by the value of the cutoff, as discussed in \cite{Greiner2}, where the
authors study the influence of the ultraviolet cutoff on the stability of cold nuclear
matter.

Nevertheless, the usefulness of NJL-type models to explore a variety of problems is well
recognized, and improvements have been achieved, in particular, concerning the
regularization procedure. In fact,  while the use of a constant cutoff in all  integrals
was a standard procedure in the former applications of the model, in the past few years
some authors have regularized the  action in order to eliminate logarithmic or quadratic
divergences only, which means that there is no need to cut the convergent integrals.
This  approach has been used   in the  evaluation of
integrals associated to triangle or box diagrams, as in \cite{Oka,CostaD} to calculate
the anomalous decays of  $\eta$ and $\pi^0$ mesons and in \cite{Jaminon} to calculate
the $\rho$ meson form factor and $\pi\pi$ scattering lengths; in both cases a better
agreement with experimental results was obtained. Recently, a regularization method that
consists in regularizing, even in the logarithmic and divergent integrals, only the
divergent parts  has been performed \cite{Oertel}.

A similar approach is nowadays used at  finite temperature, since some  integrals,
divergent in the vacuum, become convergent due to the presence of the Fermi functions;
therefore, the ultraviolet cutoff $\Lambda$ is used only in the divergent  integrals
and $\Lambda \rightarrow \infty$ in the convergent ones. This procedure was shown to have
advantages  in the study  of several thermodynamic properties
\cite{Klevansky,Greiner1,Ratti,Sasaki}.
However, the influence of this type of regularization (which from now on we denote as
regularization I) on a crucial question such as  restoration of symmetries has not yet
been analyzed. This is the main goal of the present work.

We perform our calculations in the framework of the  three--flavor  NJL model whose
Lagrangian includes the determinantal 't Hooft interaction that breaks the U$_A$(1)
symmetry:
%
\begin{eqnarray} \label{lagr}
{\mathcal L} &=& \bar{q} \left( i \partial \cdot \gamma - \hat{m} \right) q
+ \frac{g_S}{2} \sum_{a=0}^{8} \Bigl[ \left( \bar{q} \lambda^a q \right)^2+ \left(
\bar{q} (i \gamma_5)\lambda^a q \right)^2
 \Bigr] \nonumber \\
&+& g_D \Bigl[ \mbox{det}\bigl[ \bar{q} (1+\gamma_5) q \bigr]
  +  \mbox{det}\bigl[ \bar{q} (1-\gamma_5) q \bigr]\Bigr] \, .
\end{eqnarray}
%
Here $q = (u,d,s)$ is the quark field with $N_f=3$  and $N_c=3$,
$\hat{m}=\mbox{diag}(m_u,m_d,m_s)$ is the current quark mass matrix, and $\lambda^a$ are
the Gell--Mann matrices, a = $0,1,\ldots , 8$, ${ \lambda^0=\sqrt{\frac{2}{3}} \, {\bf
I}}$. The model is fixed by the coupling constants $g_S$ and $g_D$, the cutoff parameter
$\Lambda$, which regularizes the divergent integrals, and the current quark masses
$m_i\,(i=u,d,s)$. We use the parameter set \cite{Costa:2003} $m_u = m_d = 5.5$ MeV, $m_s
= 140.7$ MeV, $g_S \Lambda^2 = 3.67$, $g_D \Lambda^5 = -12.36$, and $\Lambda = 602.3$ MeV,
which is fixed by mesonic spectroscopy data: $f_\pi=92.4$ MeV, $M_\pi=135.0$ MeV,
$M_K=497.7$ MeV, and $M_{\eta^\prime}=960.8$ MeV.

The  constituent quark masses  are fixed in the vacuum by fitting the parameters of the
model to physical observables, and, in hot and dense matter, these masses depend on
temperature and density/chemical potential.
A drawback of the NJL model with the former regularization (ultraviolet cutoff in all of the
integrals {---} regularization II) is that, while, in the chiral limit, the constituent
quark masses vanish  at a critical temperature (density), away from this limit, the
masses, although decreasing, never reach its current values.
This means that chiral
symmetry  is always only approximately  restored, since the quark condensates, the order
parameters associated to the dynamical chiral  symmetry breaking, never vanish. As a
matter of fact, the constituent masses of non strange quarks go asymptotically to their
current values, and the condensates become very small at high values of temperature
(density), but the strange quark mass is always far from its current value, so it is hard
to talk about of restoration of chiral symmetry, even partial, in the strange sector.
Therefore, the use of a regularization that leads all of the constituent quark masses to
their current values could have important consequences for the restoration of symmetries,
especially the axial symmetry that we have shown to be particularly sensitive to the
behavior of the strange quark mass \cite{Costa:2004,Costa:2005,Ruivo}.

The axial U$_A$(1) symmetry is explicitly broken at the quantum level by the axial
anomaly that may be described at the semiclassical level by instantons. This  effect is
enough to generate a mass for  the $\eta'$ in the chiral limit, so this meson can not be
a remnant of a Goldstone boson  in the real world. The U$_A$(1) anomaly is responsible
for the flavor mixing effect that lifts the degeneracy between several mesons. Therefore,
if there occurs restoration of U$_A$(1) symmetry, the behavior of meson masses and mixing
angles should exhibit signals for this restoration. Another observable relevant in this
concern is the topological susceptibility $\chi$ and its slope
\cite{Meggiolaro:1992,Schafner:2000,Fukushima:2001,Costa:2004}. The topological
susceptibility is defined as
%
\begin{equation}
 \chi=\int{\rm d}^4x\;\langle
 T \{Q(x)Q(0)\}\rangle,
\end{equation}
%
where $Q(x)$ is the topological charge density. We remark that $\chi$ is related to the
$\eta'$ mass through the Witten-Veneziano formula \cite{Veneziano}. Since large instanton
effects are supposed to be suppressed at high temperatures or densities,  and
interactions between instantons contribute to the elimination of fluctuations of the topological
charge, there are good reasons to expect that the U$_A$(1) symmetry might be restored
\cite{Schaefer:2004}.

Several lattice calculations (see \cite{Alles:1997} and references therein) indicate a
sharp decrease of the topological susceptibility with temperature at zero density, and,
more recently \cite{Alles:2006}, it was shown that, at a fixed  $T$ and by varying the
chemical potential $\mu$, a critical $\mu$ is found, where the quark condensate and the
topological susceptibility drop and  the Polyakov loop raises; its derivatives   vary
sharply.

A long-standing question, for which there is yet no  answer, is whether the axial
symmetry is restored at a temperature higher than the critical temperature for the phase 
transition (scenario 1) or at about that temperature (scenario 2) 
\cite{Shuryak:1994,Pisarski:1984}, the two scenarios leading to different predictions 
concerning the behavior of chiral partners in the critical region. 
In other words, the question may be formulated as: is the restoration of chiral 
symmetry driven by quark condensates or instantons? If it is driven
by instantons, scenario 2 would be likely, implying that around  the critical point
signals of the restoration of  U$_A$(1) symmetry should already be present, as, for
instance, large fluctuations in the $\eta$ spectrum.

In order to simulate the fate of the anomaly, it is  usually assumed that the anomaly
coefficient $g_D$ is a dropping function of temperature, whether the approach is
phenomenological \cite{Alkofer:1989} or lattice-inspired
\cite{Schafner:2000,Fukushima:2001,Costa:2004}. In previous works
\cite{Costa:2004,Costa:2005}, we studied the possible effective restoration of the axial
symmetry. For the case of  finite temperature and zero chemical potential, we explored
the effects of an anomaly coupling temperature dependence $g_D (T)$ using two different
ansatz. When we modeled  $g_D (T)$  from lattice results for $\chi$, as a Fermi function,
we   found  that: (i) at $T \simeq 250$ MeV the  chiral partners ($\pi^0,\sigma$) as well
as ($a_0,\eta$) become degenerate, which is a manifestation of the effective restoration
of  chiral symmetry\footnote{The inflection point for the quark condensates, in the $T-
\mu$ plane, is taken as  the critical point for the phase transition associated with
partial restoration of chiral symmetry;  the point where the masses of chiral partners
become degenerate signals the  effective restoration of chiral or axial symmetries.};
(ii) at $T\simeq 350$ MeV,  $\chi \,\rightarrow\,0$, the pair ($\pi^0$, $\sigma$) becomes
degenerate with ($a_0$,  $\eta$) and  the mixing angles get close to the ideal values,
indicating an effective restoration of the axial symmetry. Chiral symmetry is effectively
restored before  axial symmetry, but, as shown in \cite{Costa:2005}, both symmetries can
be restored at the same point if $g_D (T)$ is chosen as an appropriate decreasing
exponential, a choice based on the phenomenological argument that  high temperature
suppresses large instanton fluctuations.
To model such a decreasing exponential,  we endow  the anomaly coefficient  with a
temperature dependence in the form \cite{Alkofer:1989}:
\begin{equation} \label{can}
g_D(T)= g_D(0)\mbox{exp}[-(T/T_0)^2],
\end{equation}
where $T_0\simeq 100 - 200$ MeV.

We verified that the whole U(3)$\otimes$U(3) symmetry is not effectively restored, 
in the range of temperatures considered for both lattice-inspired or decreasing 
exponential ansatz: (i) the strange quark condensate decreases slowly, and chiral symmetry 
in the strange sector remains broken; (ii) the mesons $\eta'$ and $f_0$ become 
purely strange, and their masses decrease moderately but do not show a tendency to converge 
or to get close to the other meson masses.

More recently, we performed a study of the relevant observables  as functions of
$\mu$ for a fixed temperature, and  we verified that the combined effect of finite $T$ and
$\mu$ does not change the usual results \cite{Ruivo}: chiral symmetry is effectively
restored only in the non strange sector, and restoration of axial symmetry is not
achieved, unless the anomaly coefficient is chosen as a dropping function of temperature
or chemical potential.

In summary, neither the ansatz used for $g_D(T)$ nor the combined effects of
temperature and density, lead to a restoration of chiral and axial symmetries in the
strange sector, unless the unrealistic condition
of equal current quark  masses  for all quarks $m_q=m_s=5.5$ MeV is imposed from 
the beginning \cite{Ruivo}.

A subject that deserves attention is, therefore, the role played by the strange quark
regarding the restoration of chiral and axial symmetries and whether the restoration of
the singlet chiral symmetry could be just  a consequence of the restoration of chiral
symmetry  or not, i.e., whether the fate of instantons drives the mechanism of
restoration of chiral symmetry or the opposite.
Moreover, the possible changes that could be induced in the scenarios described above by
allowing quark states of high momentum  to be present at high temperature should be
investigated. This is achieved  by means of  a regularization
that consists, as already referred,   in letting the  ultraviolet three-momentum cutoff  
go to infinity in all of the convergent integrals (regularization I). 
Two situations, which are summarized in Table I, will be analyzed: the anomaly coefficient 
$g_D = g_D(0)$ is kept constant, meaning that the restoration of symmetry is driven by 
the decrease of the quark condensates (Case A); in order to investigate the effect of a 
competitive mechanism, simulating the suppression of instantons at finite temperature, 
we endow  the anomaly coefficient  with a temperature dependence in the form of 
Eq. (\ref{can}), with an appropriate choice of $T_0$ (Case B). 
Both results will be compared with those obtained by regularizing all of the integrals with 
the cutoff $\Lambda$ (regularization II).
The effects of  regularization on important observables, such as the pressure and energy,
will also be analyzed.

\begin{table}[t]
\begin{center}%
\begin{tabular}
[c]{| c| c| c|}
\hline\hline
        & Anomaly coefficient $g_D$ & Regularization  \\
\hline\hline
{\bf Case A-I } & $g_D  (0)$        & $\Lambda\rightarrow \infty$    \\
{\bf Case A-II} & $g_D  (0)$        &   $\Lambda=$ Constant  \\
{\bf Case B-I } & $g_D(T)$   & $\Lambda\rightarrow \infty$    \\
{\bf Case B-II} & $g_D(T)$   &  $\Lambda=$ Constant  \\
\hline\hline
\end{tabular}
\caption{Different schemes of explicit axial symmetry breaking  and regularizations. The
anomaly coefficient $g_D(T)$ is given by Eq. (\ref{can}).}
\end{center}
\par
\label{tabelacas}
\end{table}

A remark should be added concerning the choice made for  $g_D(T)$ given by Eq. (\ref{can}).
We do not know which pattern of axial symmetry
restoration is chosen by nature, but the variation of $g_D(T)$ should not be at all
arbitrary. By using the decreasing exponential, we might choose $T_0$ in order to have two  
scenarios: (i) restoration of axial symmetry  close  to
chiral symmetry, provided $T_0$ is small enough (around $100$ MeV) \cite{Costa:2005} {---} 
this would give a critical temperature for the phase transition of about $T_c\simeq 135$ MeV, 
much lower than the present accepted values {---} and (ii) to choose $T_0$ in order to have a critical
temperature within the interval of accepted values ($T_0=170$ MeV leads to $T_c \simeq
154-163$ MeV). The last point of view will be followed here.


\section{Numerical results}

The calculations are done in a standard way \cite{Costa:2003}. With the help of the
bosonization procedure, an effective action is obtained, allowing us to evaluate a gap
equation for the constituent quark masses, the quark condensates, and the scalar and
pseudoscalar mesons. To begin with, we analyze the results for the quark masses and quark
condensates that are plotted in Fig. 1. The results with regularization II, whether $g_D$
is constant (Case A-II) or a decreasing function of the temperature (Case B-II), exhibit
the following effects:
the non strange quark masses decrease, although never attaining the current values, and
the condensates also decrease but never vanish; concerning the strange quarks, the mass
and condensate decrease moderately and are always far from the current value for the mass
and zero for the quark condensate.

\begin{figure}[t]
\begin{center}
  \begin{tabular}{ccc}
    \hspace*{-0.5cm}\epsfig{file=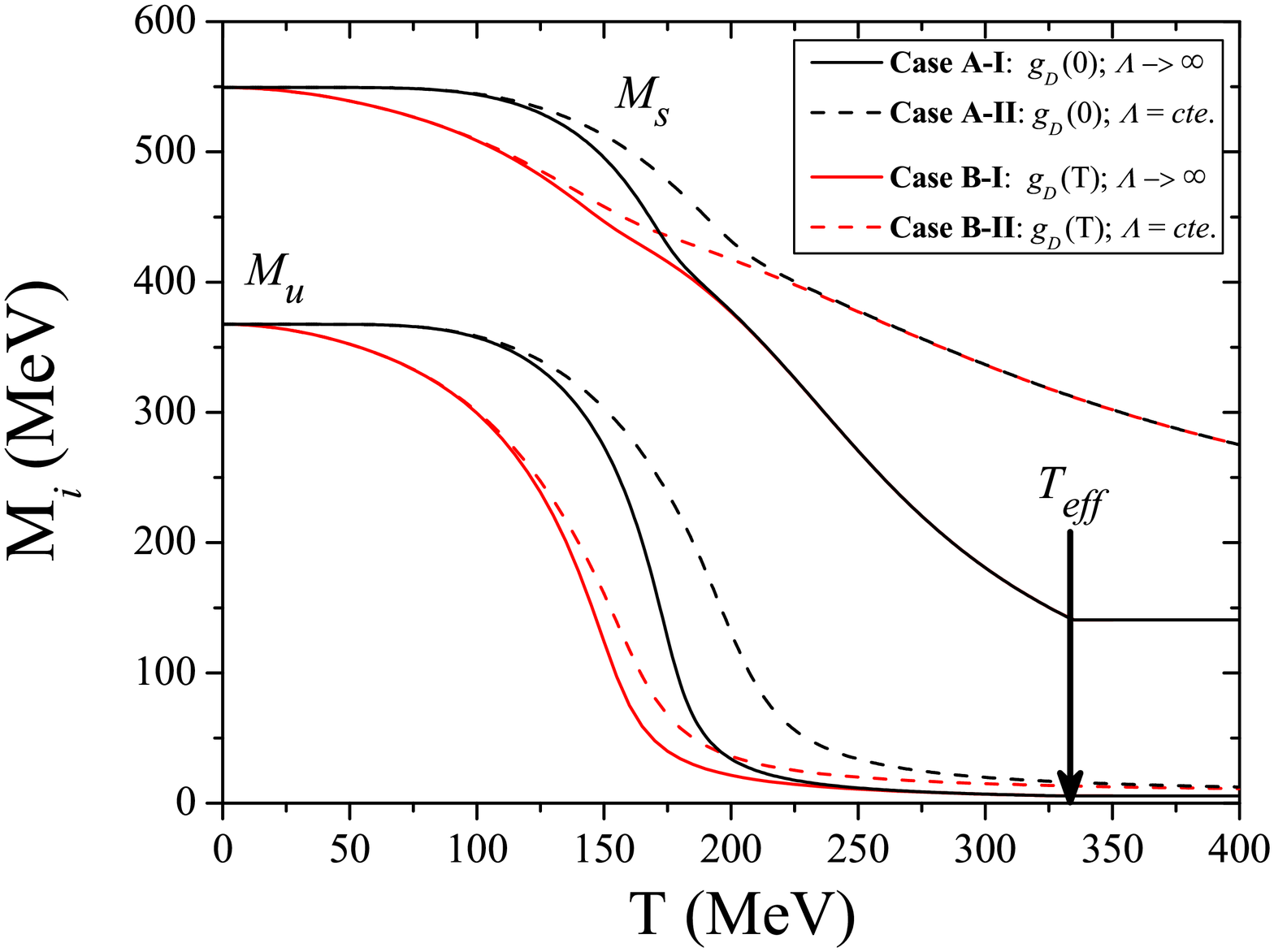,width=8cm,height=7cm} &
    \hspace*{-0.5cm}\epsfig{file=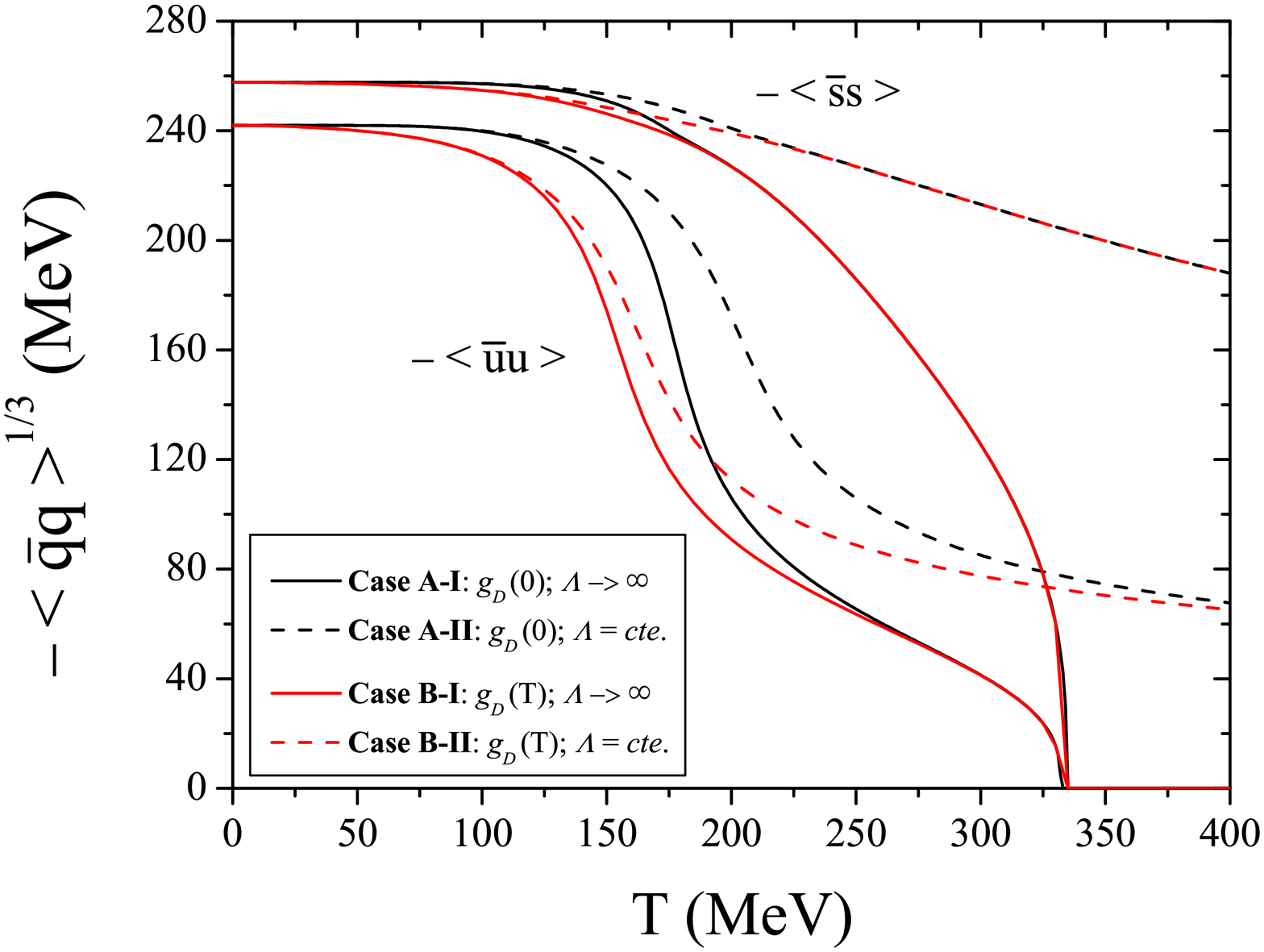,width=8cm,height=7.25cm} \\
   \end{tabular}
\end{center}
\vspace{-1.0cm} 
\caption{ Quark masses (left) and quark condensates (right)  as functions
of the temperature with two ansatz for $g_D$ and two different regularizations (Cases A
I-II and Cases B I-II.)} 
\label{fig:massescond}
\end{figure}

The situation changes drastically when we use the new regularization (Cases A-I and B-I):
all of the quark masses, at $T\simeq333$ MeV, from now on denoted as $T_{eff} $\footnote{As
matter of fact, $M_u=m_u$ for $T=333$ MeV and $M_s=m_s$ for $T=335$ MeV. Once they are
close, we adopt $T_{eff} \simeq 333$ MeV.}, go to their current values and the quark
condensates vanish, which is an indication  of the complete  restoration of the
dynamically broken chiral symmetry. This means that the contribution for the quark masses
originated by dynamical symmetry breaking completely disappears; only the current masses,
due to the explicit chiral symmetry breaking {\em ab initio}, remain.

We can also see that the
difference between the cases $g_D(0)$ and $g_D(T)$ is relevant only at low temperatures.
The new finding is that at high temperatures the dominant effect is no longer the instanton 
suppression but the decrease of quark condensates that is enhanced, especially in the 
strange sector, when high momentum quark states are allowed  ($\Lambda\rightarrow \infty$).

A remark is now in order concerning  a non-trivial  effect  of regularization I. We
notice that, above $T_{eff}$, if we do not impose any restriction, the quark masses
become lower than their current values and eventually become negative, an effect that has
been found by other authors but without its implications discussed. One can argue that
the quark masses are not observables and therefore the fact that they become negative is
not meaningful. However, when the quark masses get lower than their current values, the
quark condensates  become negative, which is not physical since they are order parameters
that should be zero in the  phase of restored chiral symmetry. Therefore, if we want to
keep calculating observables in this region, it seems sensible to impose the condition
that the quark masses take their current values and the quark condensates remain zero.
This is the approach used here.

In the left panel of Fig. 2, we plot the topological susceptibility, and we see that it
decreases  but does not vanish  when regularization II  is used; the effect of
regularization I, whether we consider $g_D (0)$ or $g_D (T)$, is the vanishing of the
topological susceptibility  at the same temperature as the quark condensates. Again we
notice the same pattern found for the masses and quark condensates: the behavior observed
at high temperatures is dominated by the effect of the infinite cutoff. Since   the
vanishing of the topological susceptibility, only by itself, does not guarantee the
restoration of the axial symmetry, we will analyze other observables.

\begin{figure}[t]
\begin{center}
  \begin{tabular}{cc}
       \hspace*{-0.5cm}\epsfig{file=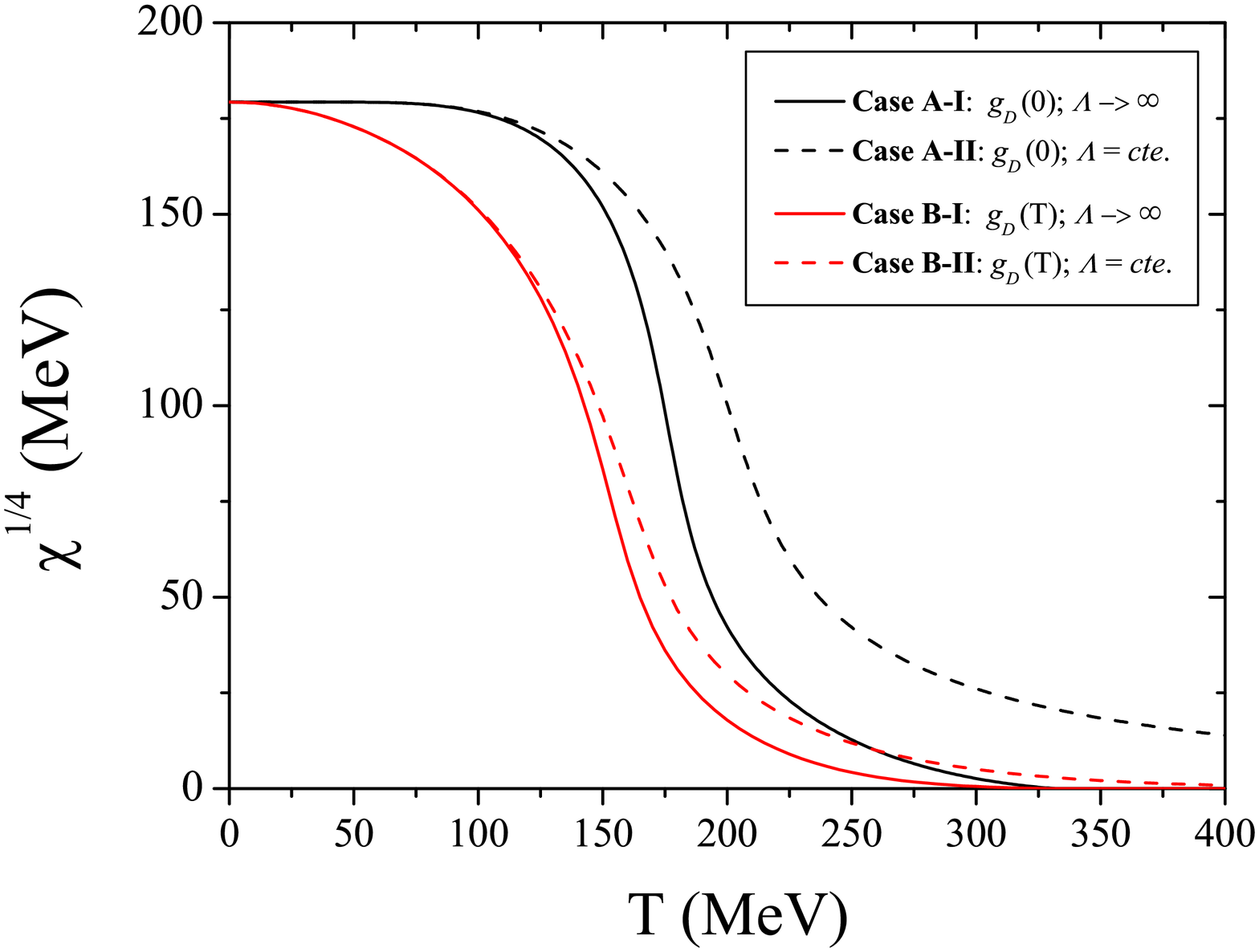,width=8.5cm,height=7cm}&
       \hspace*{-0.5cm}\epsfig{file=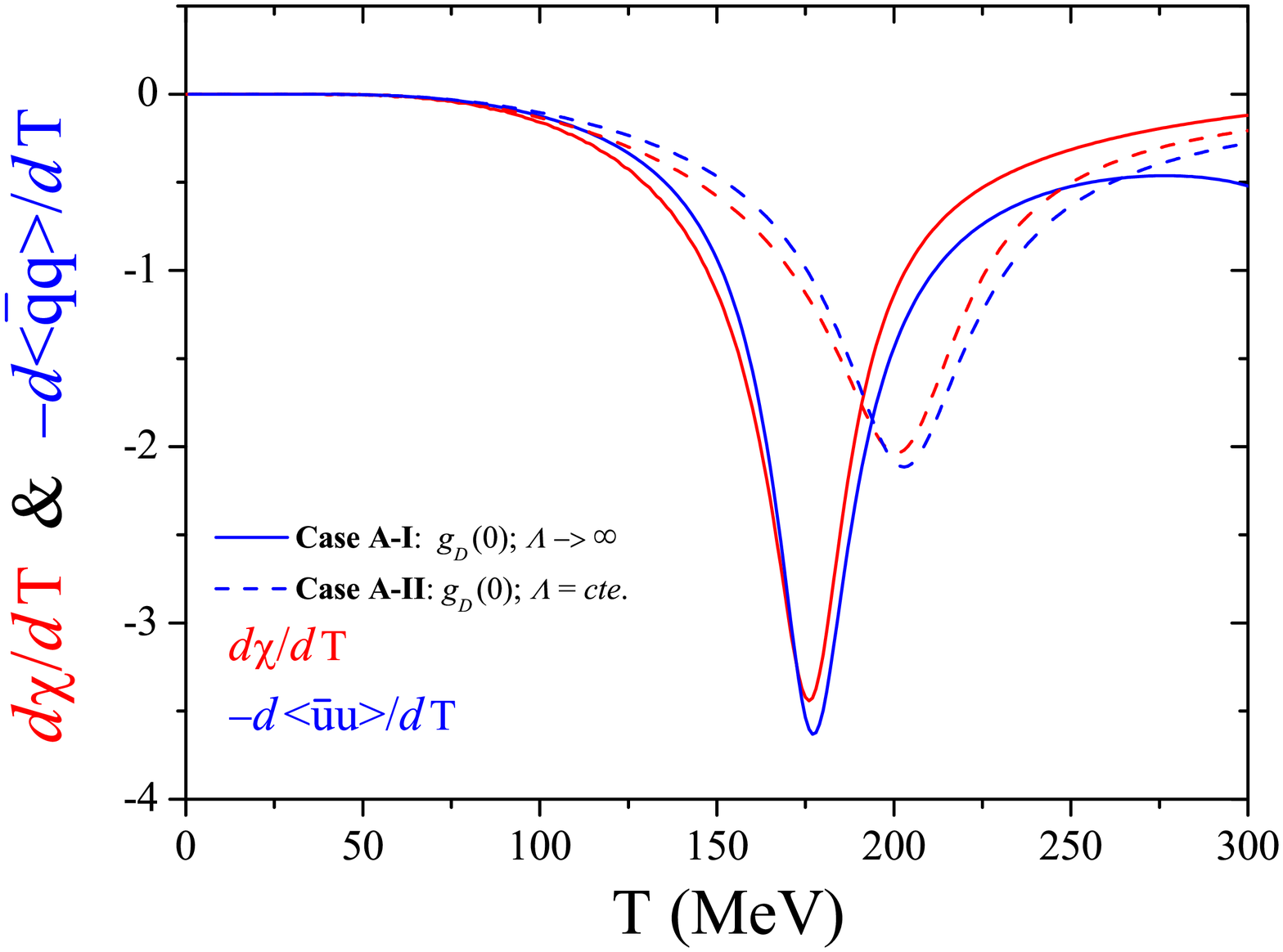,width=8.5cm,height=7cm} \\
  \end{tabular}
\end{center}
\vspace{-1.0cm} 
\caption {Topological susceptibility (left)  as function of the
temperature for Cases A I-II and  Cases B I-II. The derivatives of the quark condensates
and of the topological susceptibility (right) are shown only for Cases A I-II; a similar
pattern is found for Cases B I-II, with a shift of the inflection points for lower
temperature.}
 \label{fig:Suscep}
\end{figure}

The results of the right panel are also interesting. They show that, for both
regularizations, the inflection points of the quark condensates and topological
susceptibility occur  approximately at the same temperature, a result that has already
been found  in \cite{Alles:2006,Ruivo}; the new finding is that the critical temperature
with the new regularization is now $T_c\simeq 177$ MeV, for Case A, a value closer to the
lattice result (see \cite{lattice}) than the one obtained with regularization II
($T_c\simeq 202$ MeV). For Case B, the influence of the regularization on the value for
the critical temperature is smaller than in Case A  (see Table II).

\begin{table}[t]
\begin{center}%
\begin{tabular}
[c]{| c| c| c| c| c|}
\hline\hline
         & Case A-I & Case A-II & Case B-I  & Case B-II \\
\hline\hline
Phase Transition $(T_c)$ &  177 MeV & 202 MeV   & 154 MeV  & 163 MeV  \\
\hline
SU(2) chiral symmetry     &  200 MeV  & 250 MeV  & 180 MeV   & 205 MeV  \\
effective restoration &&&&\\
\hline
U(2) axial symmetry    &  333 MeV      &---  & 250 MeV  & 300 MeV  \\
 effective restoration   
 &&&&\\
\hline
$T_{eff}$& 333 MeV&  ---&   333 MeV & ---    \\
\hline\hline
\end{tabular}
\caption{Transition temperatures for  the different cases.  $T_{eff}$ is the transition 
temperature for the complete restoration of the dynamically broken chiral symmetry.}
\end{center}
\par
\label{tabelatemp}
\end{table}

The behavior of the mixing angles (Fig. 3) and meson masses (Fig. 4)  gives us
complementary information on the effective restoration of the symmetries under study.
Concerning the mixing angles (left panel) we see that they converge to the ideal values
in Case A-I exactly at $T=T_{eff}=333$ MeV, the temperature at which all of the quark
condensates vanish; for Case B, independently of the regularization used, the mixing
angles reach the ideal values at  lower temperatures.

\begin{figure}[t]
\begin{center}
   \includegraphics[width=0.6\textwidth]{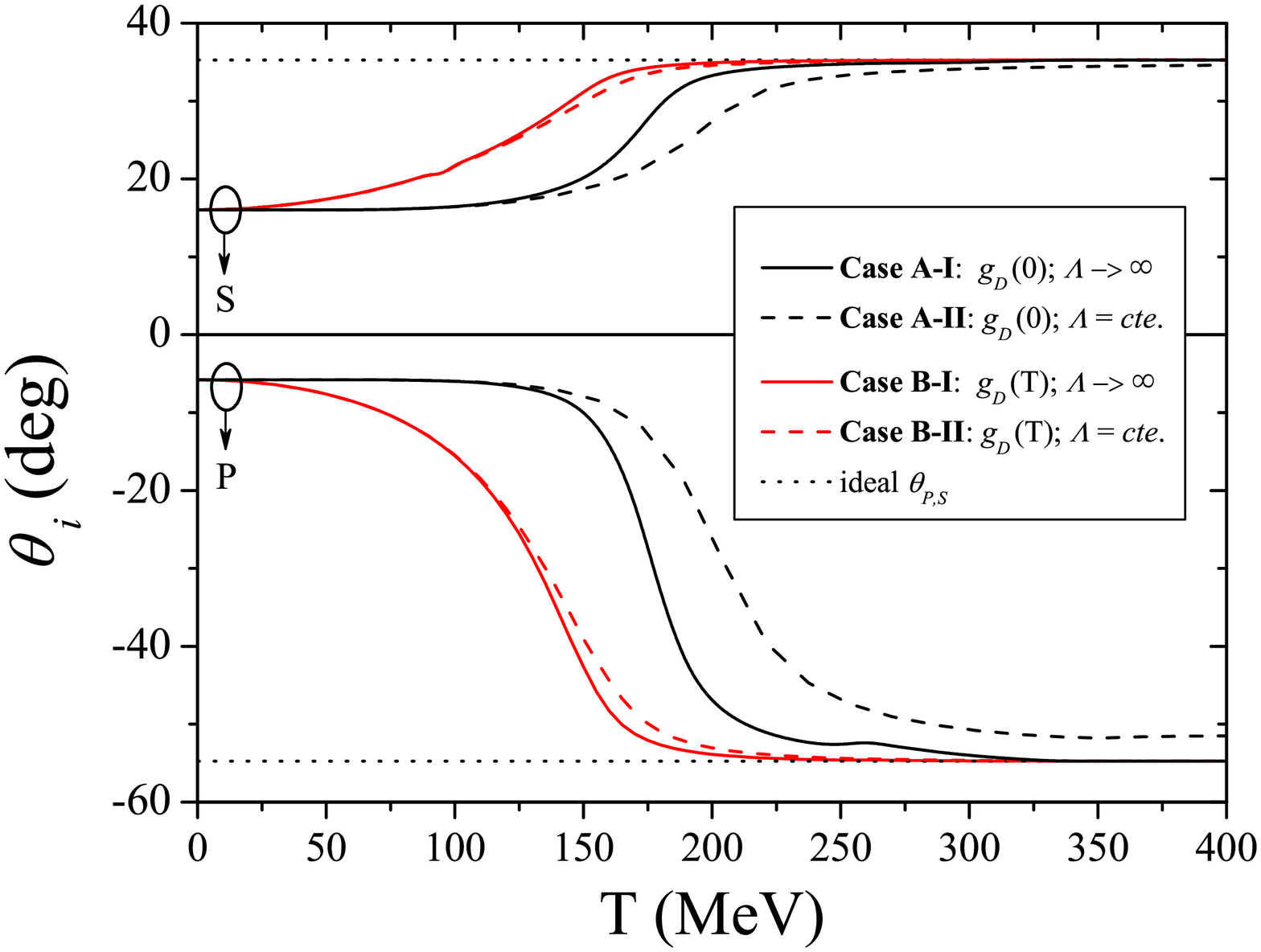}
\end{center}
\vspace{-1.0cm} 
\caption{Mixing angles  and meson masses  as functions of the temperature
for Cases A and B.} 
\label{mixing}
\end{figure}

\begin{figure}[t]
\begin{center}
   \includegraphics[width=1\textwidth]{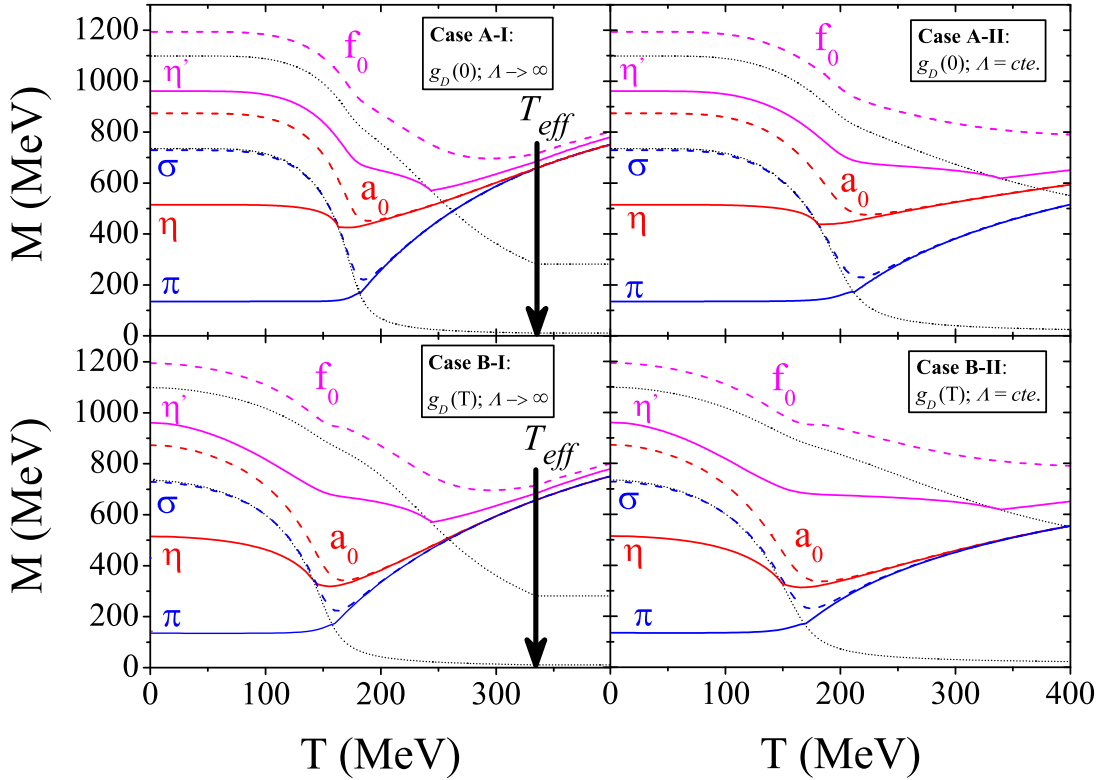}
\end{center}
\vspace{-1.0cm} 
\caption{Meson masses as functions of the temperature for Cases A and B.
The dotted lines indicate the temperature  dependence of non strange (lower curves) and
strange (upper curves) $q \bar q$ thresholds.}
\label{mesons}
\end{figure}

Prior to the analysis of  splitting between  the  masses of the chiral partners,   an
observable   that measures  the degree of effective restoration of  chiral  symmetry,
some preliminary  remarks should   be made. We notice that  most of the theoretical
insight into the problem of restoration of chiral and axial symmetries comes from lattice
calculations for a pure gauge theory and from model calculations with massless quarks or
in SU(2) models. Here we consider the physically relevant situation of explicit chiral
symmetry breaking in  SU(3) with the presence of the U$_A(1)$ anomaly, and some care
should be taken when making comparisons. The chiral partners $(\pi^0, \sigma)$ and
$(\eta, a_0)$, here studied,  have their analogs in a SU(2)$\otimes$SU(2) world without
strangeness, where the $\sigma$ and $\eta$ are non strange;  the  $(\eta', f_0)$ 
exist only in SU(3)$\otimes$SU(3). Therefore, the convergence of the two first chiral partners
is driven by the restoration of SU(2)$\otimes$SU(2) symmetry and the convergence of both
pairs, which occurs when the mixing angles go to the ideal values and all four mesons are
non-strange, indicates the effective restoration of U(2)$\otimes$U(2) symmetry. The behavior 
of $(\eta', f_0)$ is governed by the restoration of symmetry in the strange sector.

Concerning the meson masses (see Fig. 4), in Case A-I, we observe the degeneracy  of the
chiral partners $(\pi^0, \sigma)$ and $(\eta, a_0)$ at $T\simeq 200$ MeV, but both pairs
get degenerate at $T_{eff}\simeq 333$ MeV, the temperature at which the quark condensates
vanish. Comparing with Case A-II, it can be seen  (Fig. 4) that the partners $(\pi^0,
\sigma)$ and $(\eta, a_0)$ become degenerate at $T\simeq 250$ MeV (see also Table II). As
expected, the axial symmetry is not restored.

In Case B-I, the situation is qualitatively similar, but the temperatures for the phase
transition and restoration of symmetries are shifted to lower values (see Table II), and
the effective restoration of U(2) symmetry occurs at $T\simeq 250$ MeV,  which is consistent 
with the fact that the mixing angles go to the ideal values at this temperatures. The same
effect is found with regularization II (Case B-II) but at a larger temperature  
($T\simeq 300$ MeV).

Let us analyze the result with regularization I in more detail. We observe that the four
mesons ($\pi^0, \sigma, \eta, a_0$) are non strange at the temperature where the mixing
angles become ideal, even the mesons that had a component of strangeness in the vacuum,
as  $\sigma$ and $\eta$. We verified that, when the condensates are zero, the following
relations hold:
$M_{\sigma}^2\simeq4m_q^2+M_{\pi}^2$ and $M_{a_0}^2\simeq4m_q^2+M_{\eta}^2$. Since the non
strange current quark masses are negligible as compared with the meson masses at this
temperature, it is natural that  the chiral partners become  degenerated. Concerning the
behavior of the chiral partners ($f_0, \eta'$), these mesons are completely strange at
$T_{eff}\simeq 333$ MeV, and, since the strange current quark mass is high compared to
that of the non-strange quarks, it is natural that, although their masses decrease
meaningfully and get close to the other meson masses, they never converge with them. Even
$f_0$ and $\eta'$ do not become degenerate between themselves, which is due to the high
value of the current strange quark mass, since the analog of the previous relations
between meson masses is now: $M_{f_0}^2\simeq4m_s^2+M_{\eta'}^2$. In fact, the effects of
dynamical symmetry breaking vanish, in both the strange and non-strange sectors, but not
those due to explicit symmetry breaking, which are  negligible only in the non strange
sector.

A question that could be raised now is if our results fit in one of the scenarios
proposed by Shuryak \cite{Shuryak:1994} and in which of them [we recall that Shuryak's
discussion was restricted to SU(2)$\otimes$SU(2)]. Concerning the restoration of chiral
SU(2)$\otimes$SU(2) and U(2)$\otimes$U(2) symmetries, the new regularization (Case A-I)
does not lead to a change of scenario with regards to the old regularization;  however,
considering the SU(3) sector, one can say that  when high momentum quark states are
allowed we have a scenario where the axial and chiral symmetry are effectively restored
at the same temperatures $T=T_{eff}\simeq 333$ MeV.  Nevertheless, we think that the
relevant question concerning the two scenarios is not the classification but rather
whether the restoration of chiral symmetry is driven by instantons or not; in the first
case, signals of the restoration of axial symmetry should be observed at temperatures
around the critical temperature for the phase transition ($T_c$). This is not observed in
any of the cases studied here. If we take into account a mechanism simulating instanton
suppression (Case B), the restoration of axial symmetry, although taking place after the
restoration of chiral symmetry, is shifted to lower values ($T\simeq 250$ MeV, instead of
$T\simeq 333$ MeV), but this mechanism is dominant only at low temperatures (where the
infinite cutoff effect is not relevant) and in the non-strange sector.

Finally, we check the usefulness of the present regularization by plotting the energy and
the pressure as  functions of the temperature (Fig. 5), which shows that the present
results improve significantly with regards to those obtained with regularization II. So,
although not reproducing the lattice results, like the PNJL model
\cite{Ratti,Megias:2006PRD}, they are interesting from a qualitative point of view. In
fact, our results follow the expected tendency and go to the free gas (Stefan-Boltzmann)
values \cite{lattice1}, a feature that was also found in other observables with this type
of regularization \cite{Klevansky}.

\begin{figure}[t]
\begin{center}
  \begin{tabular}{cc}
    \hspace*{-0.5cm}\epsfig{file=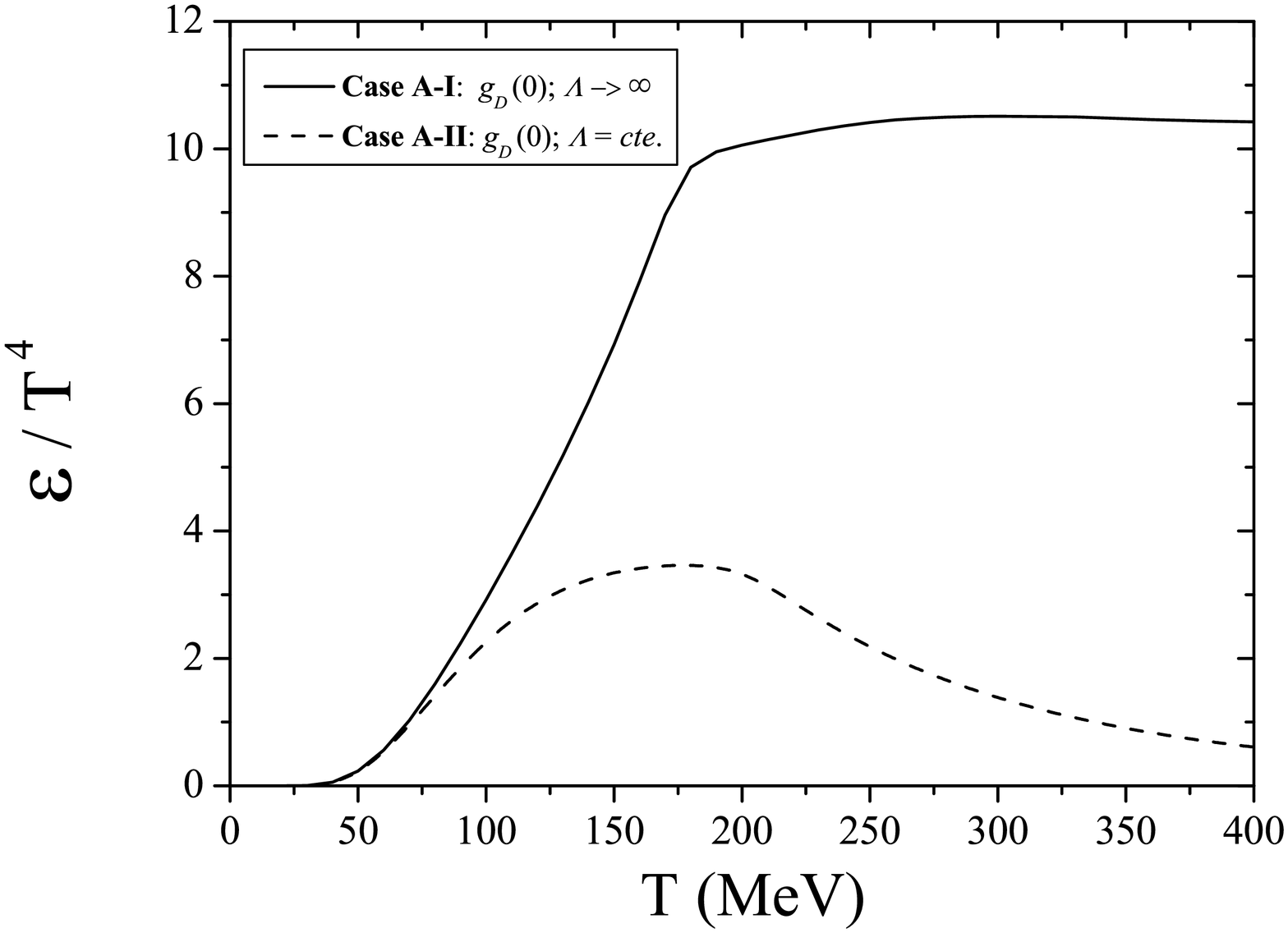,width=8.5cm,height=7cm} &
    \hspace*{-0.5cm}\epsfig{file=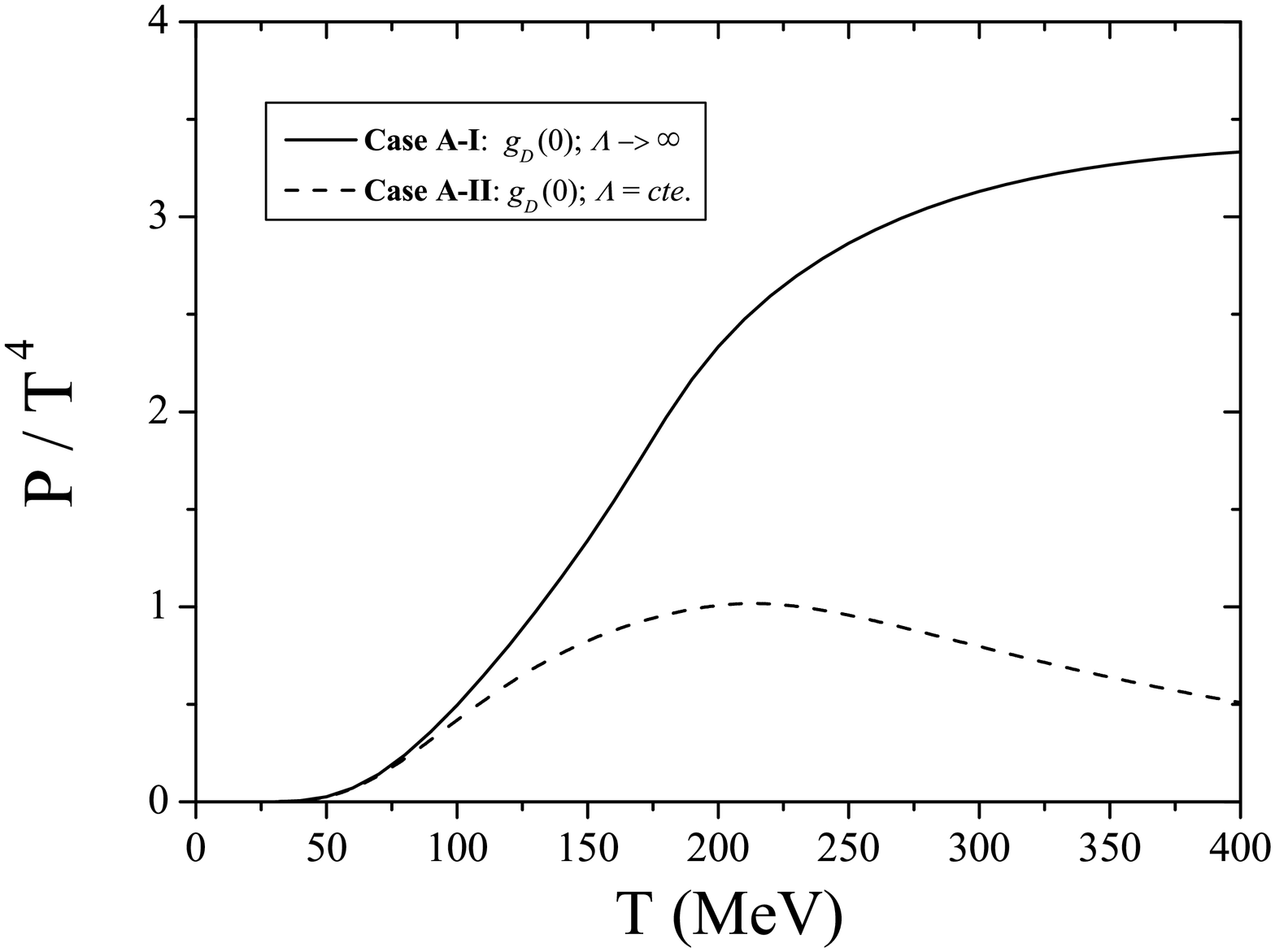,width=8.5cm,height=7cm} \\
   \end{tabular}
\end{center}
\vspace{-1.0cm} \caption{Energy and pressure as functions of the temperature for Cases A
I-II.} \label{fig:EnergPress}
\end{figure}


\section{Summary and conclusions}
 We have studied the effects on the restoration of chiral and axial symmetries of a
regularization that consists in using an infinite  cutoff in the integrals that are 
convergent at finite temperature. When the decrease of the quark condensates is  the dominant
mechanism, we found that the critical temperature, signaling the phase transition
associated with partial restoration of chiral symmetry, is lower  than with the
conventional regularization and closer to the lattice results. The main finding is  that,
with the implementation of the new cutoff procedure, restoration of chiral and axial
symmetries  can also be a phenomenon relevant in the strange sector. In fact, the 
dynamically broken chiral symmetry is completely recovered, in both the strange and non-strange sectors, 
leading to the restoration of the axial symmetry   at about the same 
temperature ($T_{eff}$): the quark masses go to the current values, the quark condensates and
topological susceptibility vanish, the mixing angles go to the ideal values, and the
masses of the mesons, which are non-strange at $T_{eff}$, converge. When an ansatz  that
simulates   independent suppression of instanton effects is taken into account, it is
shown that this  mechanism is relevant only for temperatures below $T\approx 200$ MeV,
and shifts the temperatures for the phase transition as well as  effective  restoration
of symmetries in the non-strange sector to lower values.

Regularization I also gives better results in the calculation of other observables,
like the pressure and energy, that now have the expected tendency at high temperatures.
The rich pattern of
dynamical chiral symmetry breaking/restoration   here  presented  and its relevance for other 
physical situations demands  certainly  further
investigation. On the other hand, being aware of the simplicity of our approach, we think 
that a deeper look into this problems is necessary.  Work in this direction is in progress.

\vspace{0.5cm}
This work was supported by Grant No. SFRH/BPD/23252/2005 (P. Costa) and Projects No. POCI/FP/63945/2005 
and No. POCI/FP/81936/2007 from F.C.T..



\begin{thebibliography}{}

\bibitem{Ripka1}
		G. Ripka,
		Quarks Bound by Chiral Fields, Clarendon Press, Oxford, (1997).

\bibitem{Ripka2}
    M. Jaminon, G. Ripka, and P. Stassart,
    Nucl. Phys. {\bf A504} (1989) 733;
    C. Sch{\"{o}}ren, E. Ruiz Arriola, and K. Goeke,
    Nucl. Phys. {\bf A547} (1992) 612.

\bibitem{Jaminon}
    M. Jaminon, M.C. Ruivo, and C.A. de Sousa,
    Int. J. Mod. Physics {\bf A17} (2002) 4903.

\bibitem{Greiner2}
    I.N. Mishustin, L.M. Satarov, and W. Greiner,
    Phys. Rep. {\bf 391} (2004) 363.

\bibitem{Oka}
    M. Takizawa, Y. Nemoto, and M. Oka
    Phys. Rev. D {\bf 55} (1997) 4083.

\bibitem{CostaD}
    P. Costa, M. C. Ruivo, and Yu. L. Kalinovsky,
    Phys. Lett.  B {\bf 577} (2003) 129;
    Phys. Rev. C {\bf 70} (2004) 048202.

\bibitem{Oertel}
    T. Varin, D. Davesne  M. Oertel, M. Urban,
    Nucl. Phys. {\bf A791} (2007) 422.

\bibitem{Klevansky}
    P. Zhuang, J. H{\"{u}}fner, and S.P. Klevansky,
    Nucl. Phys. {\bf A576} (1994) 525.

\bibitem{Greiner1}
    I. N. Mishustin, L. M. Satarov, H. St\"{o}cker, and W. Greiner,
    Phys. Rev. C {\bf 62} (2000) 034901.

\bibitem{Ratti}
    C. Ratti, M. A. Thaler, and W. Weise,
    Phys. Rev. D {\bf 73} (2006) 014019.

\bibitem{Sasaki}
    C. Sasaki, B. Friman, and K. Redlich,
    Phys.Rev. D {\bf 75} (2007) 054026.

\bibitem{Costa:2003}
    P. Costa, C. A. de Sousa, M. C. Ruivo, and Yu. L. Kalinovsky,
    Phys. Rev. C {\bf 70} (2004) 025204;
    Phys. Lett. B {\bf 647} (2007) 431.

\bibitem{Costa:2004}
    P. Costa, M. C. Ruivo, C. A. de Sousa, and Yu. L. Kalinovsky,
    Phys. Rev. D {\bf 70} (2004) 116013.

\bibitem{Costa:2005}
    P. Costa, M. C. Ruivo, C. A. de Sousa, and Yu. L. Kalinovsky
    Phys. Rev. D {\bf 71} (2005) 116002.

\bibitem{Ruivo}
    M. C. Ruivo, P. Costa, C. A. de Sousa
    Eur. Phys. J. {\bf A31} (2007) 766.

\bibitem{Meggiolaro:1992}
    A. Di Giacomo, E. Meggiolaro, and H. Panagopoulos,
    Phys. Lett. B {\bf 277}  (1992) 491.

\bibitem{Schafner:2000}
    J. Schaffner-Bielich,
    Phys. Rev. Lett. {\bf 84} (2000) 3261.

\bibitem{Fukushima:2001}
    K. Fukushina, K. Ohnishi, and K. Ohta,
    Phys. Rev. C  {\bf 63} (2001) 045203.

\bibitem {Veneziano}
    E. Witten,
    Nucl. Phys. B \textbf{156}, 269 (1979);
    G. Veneziano,
    Nucl. Phys. B \textbf{159}, 213 (1979).

\bibitem{Schaefer:2004}
    T. Schaefer,
    hep-ph/0412215.

\bibitem{Alles:1997}
    B. All\'{e}s, M. D Elia, and A. Di Giacomo,
    Nucl. Phys. B {\bf 494} (1997) 281.

\bibitem{Alles:2006}
    B.  All\'{e}s, M. D Elia, and M. P. Lombardo,
    Nucl. Phys. B {\bf 752} (2006) 124.

\bibitem{Shuryak:1994}
    E. Shuryak,
    Comments Nucl. Part. Phys. {\bf 21}  (1994) 235.

\bibitem{Pisarski:1984}
    R. D. Pisarski and F. Wilczek
    Phys. Rev D {\bf 29}  (1984) 338.

\bibitem{Alkofer:1989}
    R. Alkofer, P. A. Amundsen, and H. Reinhardt,
    Phys. Lett. B {\bf 218} (1989) 75;
    T. Kunihiro,
    Phys. Lett. B {\bf 219} (1989) 363;
    Z. Huang and X-N. Wang,
    Phys. Rev. D {\bf 53} (1996) 50304.

\bibitem{lattice}
    C.DeTar, et al.,
    PoS (LAT2007) 179.

\bibitem{Megias:2006PRD}
    E. Megias, E. R. Arriola, and L.L. Salcedo,
    Phys. Rev. D {\bf 74} (2006) 065005;
    Phys. Rev. D {\bf 74} (2006) 114014.

\bibitem{lattice1}
    O. Philipsen,
    arXiv:0708.1293 [hep-lat].


\end{thebibliography}
\end{document}